\documentclass[10pt,letterpaper]{article}
\usepackage[top=0.85in,left=2.75in,footskip=0.75in]{geometry}

\usepackage{changepage}

\usepackage[utf8]{inputenc}

\usepackage{textcomp,marvosym}

\usepackage{fixltx2e}

\usepackage{amsmath,amssymb}

\usepackage{cite}

\usepackage{nameref,hyperref}

\usepackage[right]{lineno}

\usepackage{xcolor}
\usepackage{microtype}
\DisableLigatures[f]{encoding = *, family = * }

\usepackage{rotating}
\usepackage{cases}

\raggedright
\usepackage[aboveskip=1pt,labelfont=bf,labelsep=period,justification=raggedright,singlelinecheck=off]{caption}

\makeatletter
\renewcommand{\@biblabel}[1]{\quad#1.}
\makeatother

\date{}

\usepackage{lastpage,fancyhdr,graphicx}
\fancyheadoffset[L]{2.25in}
\fancyfootoffset[L]{2.25in}
\nolinenumbers

\begin{document}
\vspace*{0.35in}

\begin{flushleft}
{\Large
\textbf\newline{Understanding Human Mobility from Twitter}
}
\newline
\\
Raja Jurdak\textsuperscript{1,*},
Kun Zhao\textsuperscript{1}, 
Jiajun Liu\textsuperscript{1}, 
Maurice AbouJaoude\textsuperscript{2}, 
Mark Cameron\textsuperscript{1}, 
David Newth\textsuperscript{1}
\\
\bf{1} CSIRO, Australia
\\
\bf{2} American University of Beirut, Lebanon
\\

%
%





* E-mail: raja.jurdak@csiro.au
\end{flushleft}
\section*{Abstract}
Understanding human mobility is crucial for a broad range of applications from disease prediction to communication networks. Most efforts on studying human mobility have so far used private and low resolution data, such as call data records. Here, we propose Twitter as a proxy for human mobility, as it relies on publicly available data and provides high resolution positioning when users opt to geotag their tweets with their current location. We analyse a Twitter dataset with more than six million geotagged tweets posted in Australia, and we demonstrate that Twitter can be a reliable source for studying human mobility patterns. Our analysis shows that geotagged tweets can capture rich features of human mobility, such as the diversity of movement orbits among individuals and of movements within and between cities.   We also find that short and long-distance movers both spend most of their time in large metropolitan areas, in contrast with intermediate-distance movers movements, reflecting the impact of different modes of travel.  Our study provides solid evidence that Twitter can indeed be a useful proxy for tracking and predicting human movement.


\section*{Introduction}
Understanding individual human mobility is of fundamental importance for many applications from urban planning~\cite{Noulas2012} to human and electronic virus prediction~\cite{Balcan2009, Wang2009, Tizzoni2014} and traffic and population forecasting~\cite{Wilson2004, Treiber2013}. Recently effort has focused on the study of human mobility using new tracking technologies such as mobile phones~\cite{Gonzalez2008, Jiang2013, Wesolowski2012, Palchykov2014}, GPS~\cite{Zheng2008, Injong2011, Zhao2014}, Wifi~\cite{Chaintreau2007, Zhang2012}, and RFID devices~\cite{Cattuto2010, Fournet2014}. While these technologies have provided deep insights into human mobility dynamics, their ongoing use for tracking human mobility involves privacy concerns and data access restrictions. Additionally, the use of call data records from cellular phones to track mobility provides low resolution data typically in the order of kilometres, dictated mainly by the distances between cellular towers. 

Recently,  large online systems have been proposed as proxies for providing valuable information on human dynamics~\cite{Hawelkaa2013, Frank2013, Wu2014, Wang2014, Austin2014}. For example, the online social networking and microblogging system Twitter, which allows registered users to send and read short text messages called tweets, consists of more than 500 million users posting 340 million tweets per day. Users can opt to geotag their tweet with their current location, thus providing  an ideal data source to study human mobility.  Geotagged tweets provide high position resolution down to 10 metres together with a large sample of the population, representing a unique opportunity for studying human mobility dynamics both with high position resolution and at large spatial scales.

Despite the data being publicly available and having a large population of users, its representativeness of the underlying  mobility dynamics remain  open questions. Specifically, there are three open issues with using geotagged tweets for understanding mobility patterns: (1) potential sampling bias; (2) communication modality; and (3) location biases for sending tweets. As a social networking service, the population of Twitter users provides a specific sample of the population where people must have an Internet connection, be relatively tech savvy, and thus typically represent a younger demographic group. While sampling bias is likely to be prevalent for any technology that captures mobility dynamics~\cite{Yan2013, Wesolowski2013}, it is unclear how Twitter's potential sampling bias affects the mobility patterns of geotagged tweets.  Another challenge is that Twitter, unlike previous technologies~\cite{Zhao2011},  strictly limits content length within one message. To use Twitter as a proxy for studying human mobility, it is important to understand whether this hard limit on tweet content can impact the spatiotemporal patterns of geotagged tweets. Finally, it is currently unclear whether Twitter users send messages from specific types of locations (such as the home or workplace), and how such preferences to send tweet messages from certain locations can impact the mobility patterns observed from geo-tagged Tweets.

This paper analyses a large dataset with $7,811,004$ tweets from $156,607$ Twitter users from September 2013 to April 2014 in Australia to determine how representative are Twitter-based mobility patterns of population and individual-level movement. We compare the mobility patterns observed through Twitter with the patterns observed through other technologies, such as call data records. Our analysis uses universal indicators for characterising mobility patterns from geotagged tweets, namely the displacement distribution and gyration radius distribution that measures how far individuals typically moves (their spatial orbit). We find that the higher resolution Twitter data reveals multiple modes of human mobility~\cite{Balcan2009} from intra-site to metropolitan and inter-city movements. Our analysis of the time and likelihood of returning to previously visited locations shows that the strict content limit on tweets does not affect the returning patterns, although Twitter users exhibit higher preference than mobile phone users for returning to their most popular location. We also observe that an individual's spatial orbit strongly correlates with their mobility features, such as their likelihood and timing to return to their home location, suggesting that categorising people by their spatial orbit can improve the predictability of their movements. We notably find that short and long-distance movers both spend most of their time in large metropolitan area, in contrast with intermediate-distance movers movements, reflecting the impact of different modes of travel. In studying the predictability of next tweet location, we find evidence of two types of Twitter users, mapping to a highly predictable group that appears to have strong spatial preference for tweeting, and a less predictable group for which Twitter is a better proxy to capture mobility patterns.  

\section*{Results}

\subsection*{Displacement distribution and Radius of gyration}
We first characterise the movement patterns of individuals by analysing their sequences of geotagged tweets. Based on these sequences, we can study the moving distances for individual trips and over the long-term.

\begin{figure}[h]
    \centering
     \includegraphics[width=\columnwidth]{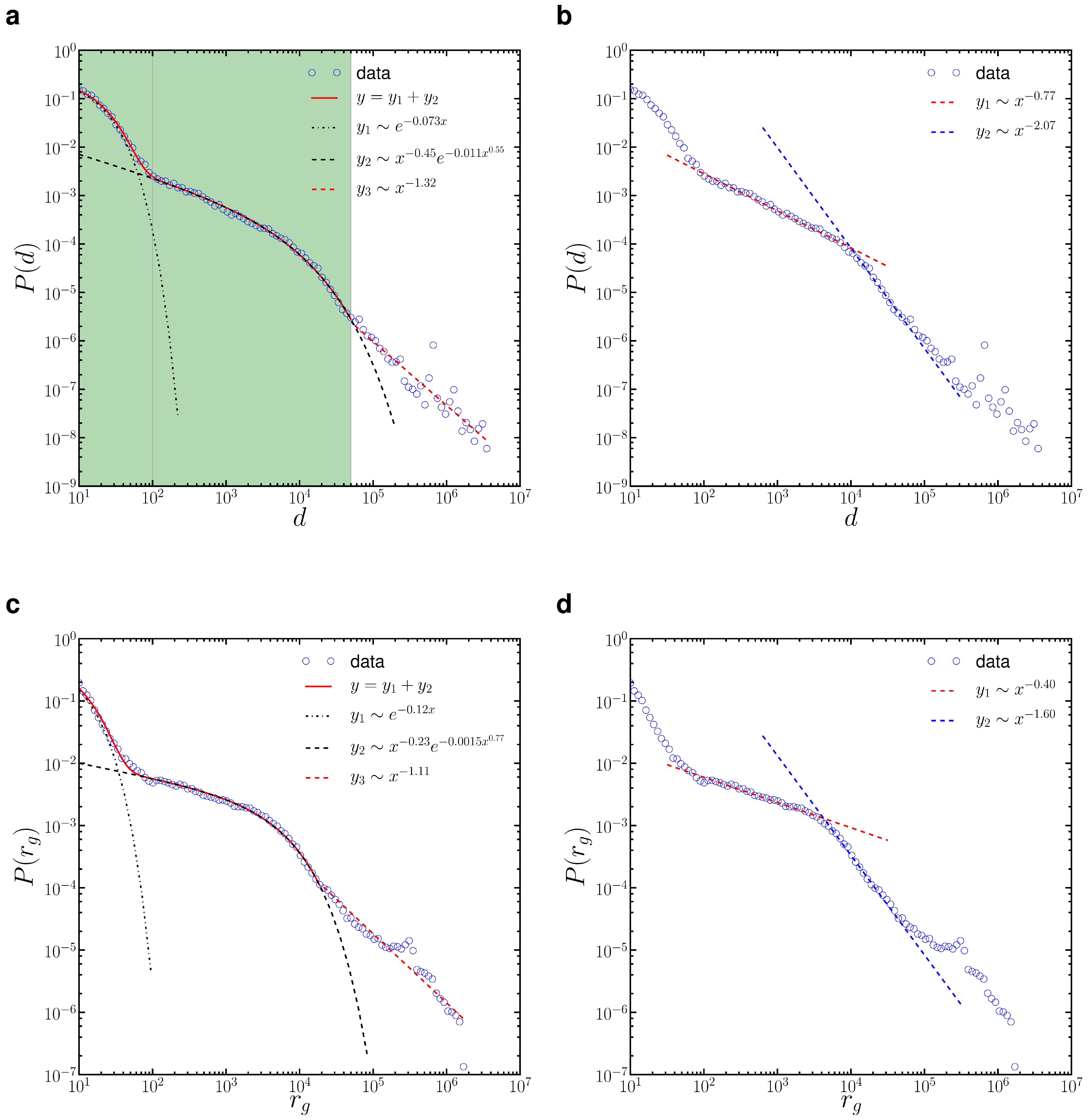}   

\caption{{\bf (a) - (b) Displacement distribution $P(d)$ observed from Twitter geotagged data with different fitting schemes.} In panel (a),  $P(d)$ is approximated by a mixture function Eq.(\ref{fit1}) adjoined with a power-law, indicating the presence of multi-modal or multi-scale mobility in Australia. In panel(b), $P(d)$ with $d \in [100m, 50km]$ is approximated by a double power-law, indicating the intra-city or urban movements may be composed of two separate modes. (c) - (d) Distribution of gyration radius $P(r_g)$ follows a remarkably similar pattern as $P(d)$, and is fitted by the two fitting schemes as done for $P(d)$. The details of these two fitting schemes and statistical validation are given in the Supporting Information S1 Text. }
\label{fig:1}
\end{figure}

The first important characteristic in human mobility patterns is the displacement distribution, namely spatial dispersal kernel $P(d)$~\cite{Brockmann2006, Gonzalez2008} , where $d$ is the distance between a user's two consecutive reported locations. Figure \ref{fig:1}(a) shows that the displacement distribution $P(d)$ is characterised by a heterogenous function with $d \in [10m,4000km]$ spanning more than five decades horizontally.  Note that the observed patterns here can be affected by the variation of location records due to the resolution limits of devices, particularly for short distance steps. Therefore, displacements shorter than $10$ meters ($d<10m$), which may represent positioning noise rather than actual displacement due to typical resolution limits of GPS devices~\cite{Jurdak13}, are not considered in this study. The results are also dependant on the tweeting dynamics that shape the temporal domain of the location records. We discuss the possible impacts from these two aspects in the Supporting Information S1 Text.

The heterogenous shape of $P(d)$ for the entire interval can be hardly captured by a single commonly-used statistical function such as a power-law or an exponential using the approach of parametric fitting. Indeed, we find that $P(d)$ can be better approximated by a combination of multiple functions, indicating the presence of multi-modality in human mobility patterns. 

One scheme of fitting $P(d)$ is to use a hybrid function 
\begin{equation}
P(d) \sim q \lambda_1 e^{-\lambda_1(d-d_{min})} + (1-q) \beta \lambda_2   d^{\beta-1} e^{-\lambda_2 (d^{\beta} - d_{min}^{\beta})}, \label{fit1}
\end{equation}
which is a mixture of two individual functions, namely an exponential and a stretched-exponential. Here $\lambda_1$, $\lambda_2$ and $\beta$ are the parameters that control the shape of each individual function, $q$ is a coefficient that controls the mixture proportion, and $d_{min} = 10m$ is the minimum bound of the distribution. We find that, Eq.(\ref{fit1}) can fit the displacement distribution perfectly up to a cut-off $d_c \approx 50km$ (the shaded area in Figure \ref{fig:1}(a)), which may reflect the range limit of daily urban mobility such as commuting or shopping. Therefore it is reasonable to argue that  this part of the distribution, taking up about $93.6\%$ of the total displacements, approximately captures the population-level trend of urban movements in Australia. The mixture function, with a significant inflection at $d_e \approx 100m$, may account for two different modalities of urban movements:  (1) intra-site movements; (2) metropolitan movements. In particular, the first exponential function, which dominates for short displacements but declines dramatically as $d$ increases, represents the intra-site movements such as short relocations within a building. Meanwhile, the stretched-exponential function that dominates for $d \ge d_e \approx 100m$ represents the metropolitan movements. The transition between these two modes at $d_e \approx 100m$ aligns well with the typical magnitude of site size. The noisy tail of the distribution beyond the urban mobility cut-off $d_c \approx 50km$, accounting for about $6.4\%$ of the total displacements, can be roughly approximated by a power-law. This part may represent the long-distance or inter-city traveling mode. The humps in this part can be attributed to the sparse and concentrated population distribution as well as the scattered distribution of major metropolitan areas in Australia. For example, the significant hump between 600km-1000km is very likely due to the frequent traveling trips between large cities in Australia. We discuss this peak in detail shortly which we attribute to intercity movement, a third mode in mobility, in Section~\ref{sec:Pxy}.   

The dominating stretched-exponential distribution for $d \in [100m, 50km]$ that accounts for most intra-city or urban movements in Australia is in contrast to the results from other studies such as the power-law from banknotes~\cite{Brockmann2006}, the truncated power-law from mobile phones~\cite{Gonzalez2008} and travelling surveys~\cite{Yan2013}, the exponential \cite{Liang2013} and more recently the log-normal \cite{Zhao2014} from high-resolution GPS tracking for various urban transportations such as taxis or cars. The unique stretched-exponential observed in our study indicates that the travelling distance of urban movements in Australia may stem from multiplicative processes, i.e. the displacement $d$ is determined by the product of $k$ random variables~\cite{Laherrere1998, Frisch1997}. These random variables can be transportation cost, lifestyle aspects such as the preference on commute distance, or socio-economic status such as personal income. The number of these variables $k$, namely the number of levels in the multiplicative cascade, is indicated by the exponent $\beta$ in Eq.(\ref{fit1}). When $k$ is small, $P(d)$ converges to a stretched-exponential asymptotically, and $k \rightarrow +\infty$ leads to the classic log-normal distribution. In particular, if these random variables are Gaussian distributed, we have $k \approx 2/ \beta$, and the value of $k$ is around $3$ or $4$ for our data ($\beta \approx 0.55$). 

While stretched-exponential is the most competitive candidate compared to other single statistical functions, an alternative fitting scheme for $d \in [100m, 50km]$ is to use a double power-law function shown in Figure~\ref{fig:1}(b):
\begin{numcases}{P(d) \sim }
  d^{-\gamma_1}  & $d_{min} \le d < d_m$ \nonumber \\
  d^{-\gamma_2}  & $d_m \le d < d_c$ \label{fit2}
\end{numcases}
where $\gamma_1$ and $\gamma_2$ are the exponents for each individual power-law and $d_m$ is the separation point. This scheme suggests that the urban mobility mode can be also comprised of two separate modes characterised by two different power-laws, possibly capturing differences between short and long distance moves within a city.

To explore the heterogeneity of mobility among individuals, we study the radius of gyration $r_g$, which is another important characteristic of human mobility that quantifies the spatial stretch of an individual trajectory or the traveling scale of an individual~\cite{Gonzalez2008, Jiang2013}. The radius of gyration for an individual can be calculated by $r_g = \sqrt{\frac{1}{n}\sum_i (\vec{r}_i - \vec{r}_c)^2}$, where $\vec{r}_i$ is the individual's i-th location, $\vec{r}_c = \frac{1}{n}\sum_{i}\vec{r}_i$ is the geometric center of the trajectory and $n$ is the number of locations in the trajectory. The distribution of the radius of gyration $r_g$ over the whole population  in Figure~\ref{fig:1}(c) has a similar shape as observed in the displacement distribution, which indicates that there is strong individual heterogeneity of traveling scale over the whole population. In other words, there may be geographical constraints that shape the movement depending on how far people move from their home location. Indeed, it has been suggested that the distribution of $r_g$ should be asymptotically equivalent to $P(d)$, if each individual trajectory is formed by displacements randomly drawn from $P(d)$ \cite{Song2010}. The general shape of the $r_g$ distribution looks similar to some previously reported $r_g$ distributions in other regions such as Santo Domingo in the Dominican Republic~\cite{Ji2011}, suggesting highly regional specific dynamics at play, such as population density and urban layout. However, our geotagged Twitter data provides position resolution at up to 10m, compared to typical resolutions of 1km in previous studies~\cite{Gonzalez2008,Song2010}, allowing more fine-grained validation of these dynamics. 

\subsection*{First-passage time and Zipf's law of visitation frequency}
\begin{figure}[h]
    \centering
\includegraphics[width=\columnwidth]{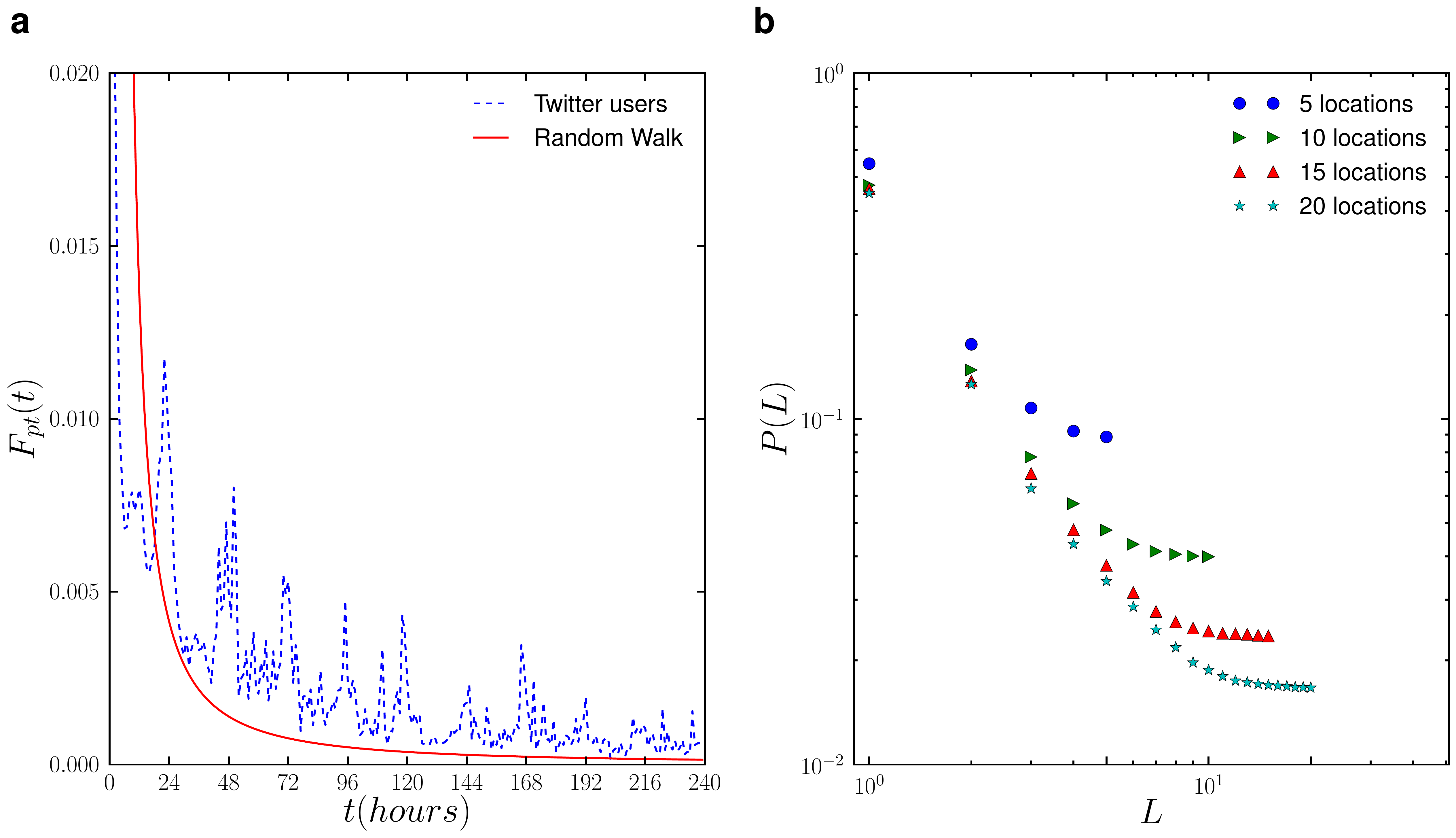}
\caption{{\bf Return dynamics of Twitter users: (a) The probability of return of users to a given locational; (b) preferential return to previously visited locations.}}
\label{fig:2}

\end{figure}

To gain better insight into the individual mobility patterns, we measure the first-passage time probability $F_{pt}(t)$, i.e. the probability of finding a user at the same location after a period of $t$, as shown in Figure \ref{fig:2}(a). We observe a periodic fluctuation at a 24-hour interval in $F_{pt}(t)$, representing the home-return patterns in human mobility, similar to call data records~\cite{Gonzalez2008}. This confirms that the periodic patterns are indeed strongly present in human movement and are independent of communication medium proxy.  The similarity of $F_{pt}(t)$ between geotagged tweets and call data records provides strong validation that the intrinsic differences between the two communication media do not significantly affect their observed return dynamics. Each tweet is a short message of up to 140 characters,  allowing people to communicate only one idea at a time. People may then send multiple consecutive tweets from the same location to convey a series of ideas. This is in contrast to phone calls  where people can convey multiple ideas within a single call without hard constraints on content volume. From the perspective of using data traces of these two technologies as proxies for human mobility, locations from consecutive geotagged tweets are thus much more likely to be the same compared to locations from consecutive phone calls. Despite this difference, the first passage time patterns are similar. 

We are also interested in how well geotagged tweets can reflect the visitation preference of locations.  We therefore measure the probability function $P(L)$ of finding an individual at his/her $L$-th most visited location. $P(L)$ can be obtained by sorting an individual's visited locations in descending order of the visitation frequency, and visited locations can be identified by performing spatial clustering with a radius of 250m on the raw data (see Methods and ~Supporting Information S1 Text). As shown in Figure \ref{fig:2}(b),  we observe a Zipf's law of the visitation frequency~\cite{Zipf1946}, i.e. $P(L)$ can be described by a power-law function $P(L) \sim L^{-\alpha}$. It has been recently suggested that the Zipf's law of the visitation frequency in human mobility is rooted in the preferential return dynamics~\cite{Song2010},  i.e. humans have a tendency to return to the locations they visited frequently before. In this context, the exponent $\alpha$ reflects the strength of the preferential return. In particular, when $\alpha = 0$, the individual returns to the visited locations with equal probability, and as $\alpha$ increases, the individual has a stronger tendency of returning to more frequently visited locations. 

Moreover, we observe that $P(L=1)$, the probability of finding an individual in his/her most frequently visited location (or home location), ranges between 0.45 to 0.55 depending on the number of visited locations, which is significantly higher than the value observed in the mobile phone records~\cite{Song2010}. This finding indicates that people are likely to tweet in their most popular or home location at nearly half the time. This consistent higher likelihood to tweet from the home location is again most likely explained by difference in technology use.  People are more likely to make a cellular phone call outside the home location, while tweeting appears to be more preferential at home. It is important that mobility patterns extracted from Twitter consider this higher preference for the home location for representative models. Other than this distinction, geotagged tweet data appears to align well with the preferential return patterns observed previously across large populations. 

\subsection*{Predictability of Individual Tweet Locations}
To better understand the individual mobility pattern and its correlation with tweeting behaviour, i.e. where and when does a user tweet, we study the randomness and predictability of the sequence of tweeting locations for each user. To this end, we measure the entropy of each individual sequence of tweeting locations, which is a common approach to capture the randomness and predictability of time-series data~\cite{Song2010_2, Lu2013}. Here we consider two typical entropy measures (see Methods for more details) : (1) the unconditional entropy $S_{unc}$ (Shannon entropy) that merely measures the randomness based on the occurring frequency of each distinct tweeting location; and (2) the real entropy $S_{real}$ that evaluates the randomness based on the full information of the sequence, i.e. not only the occurring frequency of each distinct location but also the order of locations, capturing the intrinsic spatiotemporal correlation of individual mobility as well as tweeting behaviour. In general we have $S_{unc} \le S_{real}.$

\begin{figure}[h]
\centering
\includegraphics[width=\columnwidth]{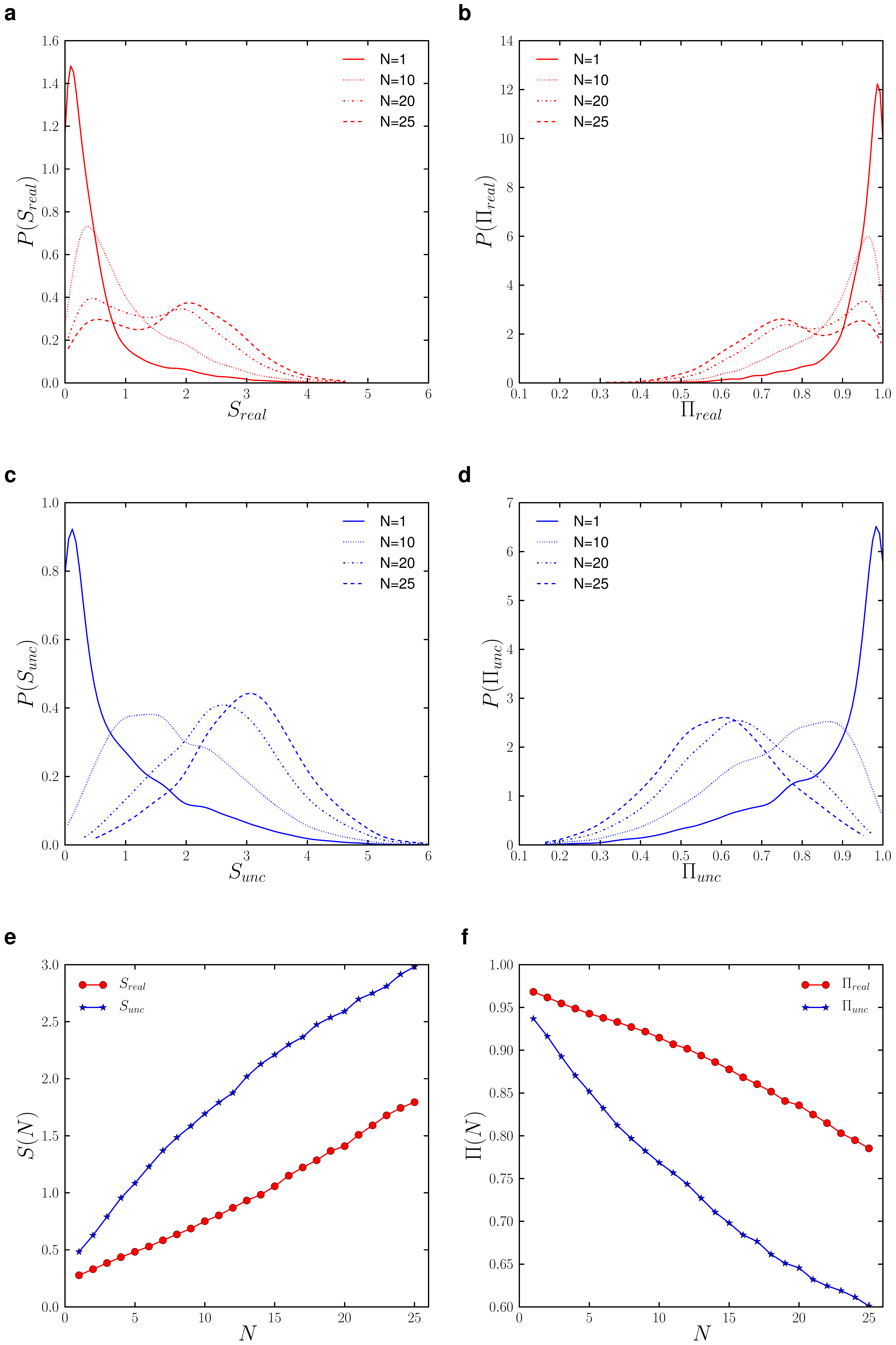}
\caption{ {\bf The entropy and predictability of individual tweet locations:} $S_{real}$ (a) and $S_{unc}$ (c)  expectedly increase with the minimum number of tweet locations. A bimodal distribution in $S_{real}$ emerges from N=20 and becomes more pronounced at N=25 pointing to a  group with very low randomness and high predictability and another group among users with highly diverse tweeting locations and lower predictability (b).  The rates of entropy increase (e) and predictability decrease (f) are significantly lower for $S_{real}$ compared to $S_{unc}$. }
\label{fig:tweet_predict}
\end{figure}

We first determine $S_{unc}$ and $S_{real}$ for users with at least $100$ tweets ($8567$ users) and study their distribution $P(S_{unc})$ and $P(S_{real})$ across the user population, with the results shown in Figure~\ref{fig:tweet_predict}. One simple quantity which is closely related to the entropy and predictability of an individual location sequence is the number of distinct locations $N$ in that sequence. The median value of both $S_{unc}$ and $S_{real}$ for users with different values of $N$ approximately follow a linearly increasing trend as $N$ increases (Panel e), as a higher diversity of tweeting locations is more likely to have a higher order of randomness as well. However, $S_{unc}$ increases at a faster rate, which indicates that the gap between the gain of randomness based on the occurring frequency and the full information enlarges as $N$ increases, i.e. in terms of reducing the uncertainty of the user's next tweeting location, the advantage of using full historical information compared to using occurring frequency only becomes more prominent for larger $N$. The distributions $P(S_{unc})$ and $P(S_{real})$ are obtained for four user groups with a minimum number of distinct tweeting locations $N = 1, 10, 20, 25$ respectively (Panel a and c). For small values of $N$ (e.g. N = 1, 10), both $P(S_{unc})$ and $P(S_{real})$ appear to be bell-shape and peak at a center value $S^{peak}$. As $N$ increases, $S^{peak}$ shifts to a larger value and the overall distribution becomes broader. The more striking result is the emergence of a bimodal distribution for $P(S_{real})$ when $N$ exceeds $N=20$, with the second peak becoming more pronounced for $N=25$. This result may suggest the existence of two distinct types of Twitter users: a group that maintains low randomness and high regularity (low entropy and high predictability) in their tweeting location patterns despite high diversity of tweeting locations; and another group that exhibits high randomness and low regularity in their tweeting location patterns which are consistent with their high diversity of tweeting locations. This bimodal distribution can be rooted in the users' real spatial mobility, or the correlation between the tweeting preference and the mobility, keeping in mind that the sequence of tweeting locations is only a subset of the user's real visited locations. In terms of the latter cause, the mobility patterns of the first group are more likely attributed to their tweeting location preference, and are not necessarily representative of their full mobility patterns. For the second group, the breadth of tweeting locations along with a comparably high entropy suggests that their tweeting locations are less likely to be correlated with the tweeting preference and therefore appear to be a more representative indicator of their real mobility patterns. 

Another important measure for predictability is the maximum bound of probability $\Pi$ (or the maximum predictability) that an appropriate prediction algorithm \cite{Gambs2012} can predict the user's next location. 
For instance, a user with a maximum predictability $\Pi=0.4$ has at most 40\% of his/her tweeting locations that can be predicted, while at least 60\% of his/her tweeting locations appear to be random and unpredictable. $\Pi$ can be estimated from the entropy using Fano's inequality (see Methods).
The trends of $\Pi_{unc}$ and $\Pi_{real}$ as a function of $N$ (Panel f) and the distributions $P(\Pi_{unc})$ and $P(\Pi_{real})$ (Panel b and d) clearly mirror the results observed in the entropy, with a lower predictability evident for a larger number of tweeting locations, and a bimodal distribution of predictability for $20$ or more tweeting locations. The decrease in predictability for higher $N$, similarly, is much slower for $\Pi_{real}$ compared to $\Pi_{unc}$, highlighting the importance of spatiotemporal correlations in visitation patterns in predicting future tweet locations. 

\subsection*{Probability density function P(x,y)}
\label{sec:Pxy}
\begin{figure}[h]
\centering
\includegraphics[width=\columnwidth]{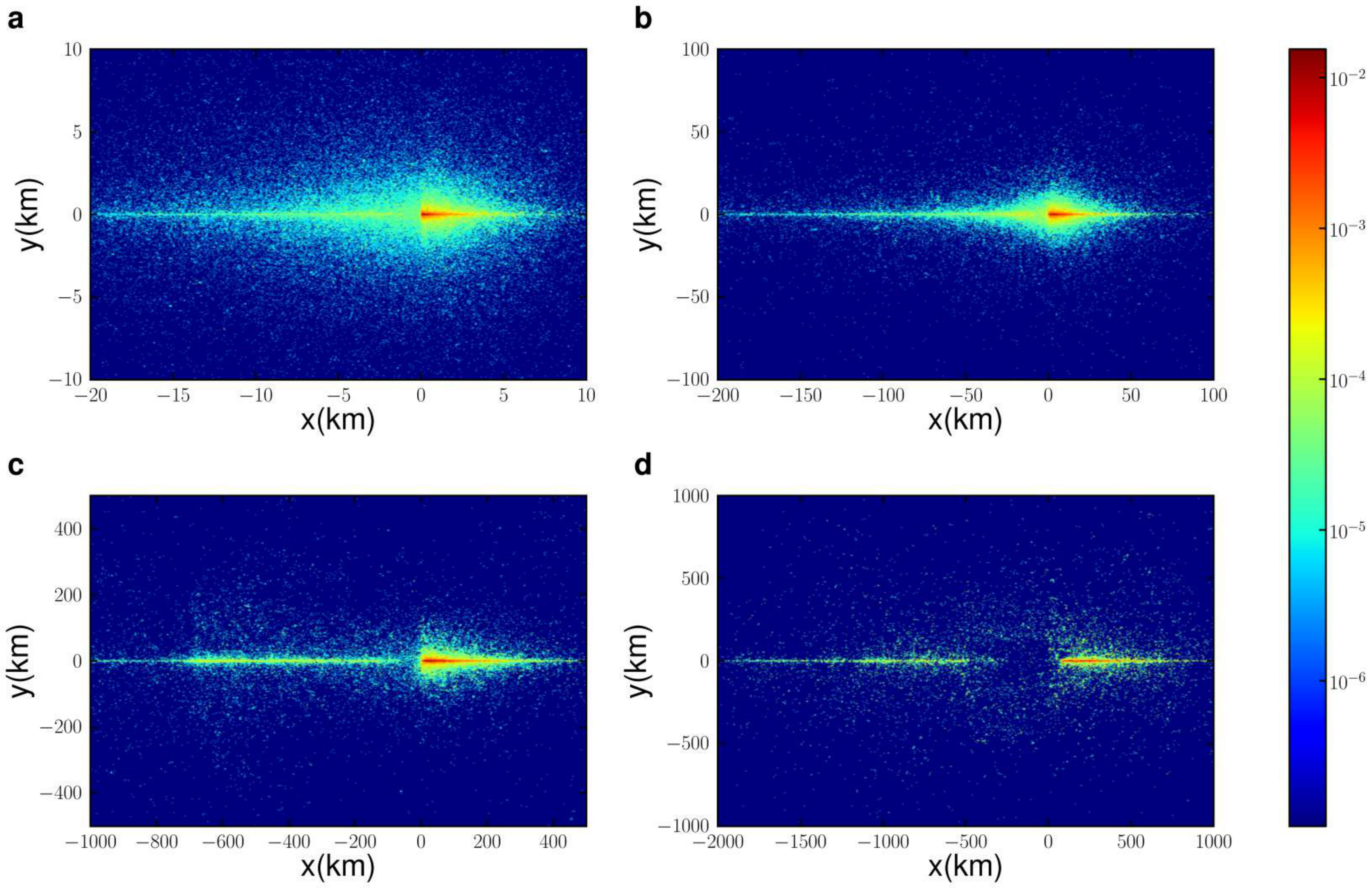}
\caption{{\bf The probability density function $P(x, y)$ for users with different radius of gyration: (a) 1-10 km; (b) 10-100 km; (c) 100-500 km; (d) $>$500 km.} Directed motion initially increases with $r_g$, mainly influenced by road usage, yet motion patterns become more diffusive for large distances with the increased use of air travel.}
\label{fig:Pxy}
\end{figure}

\begin{figure}[h]
\centering
\includegraphics[width=\columnwidth]{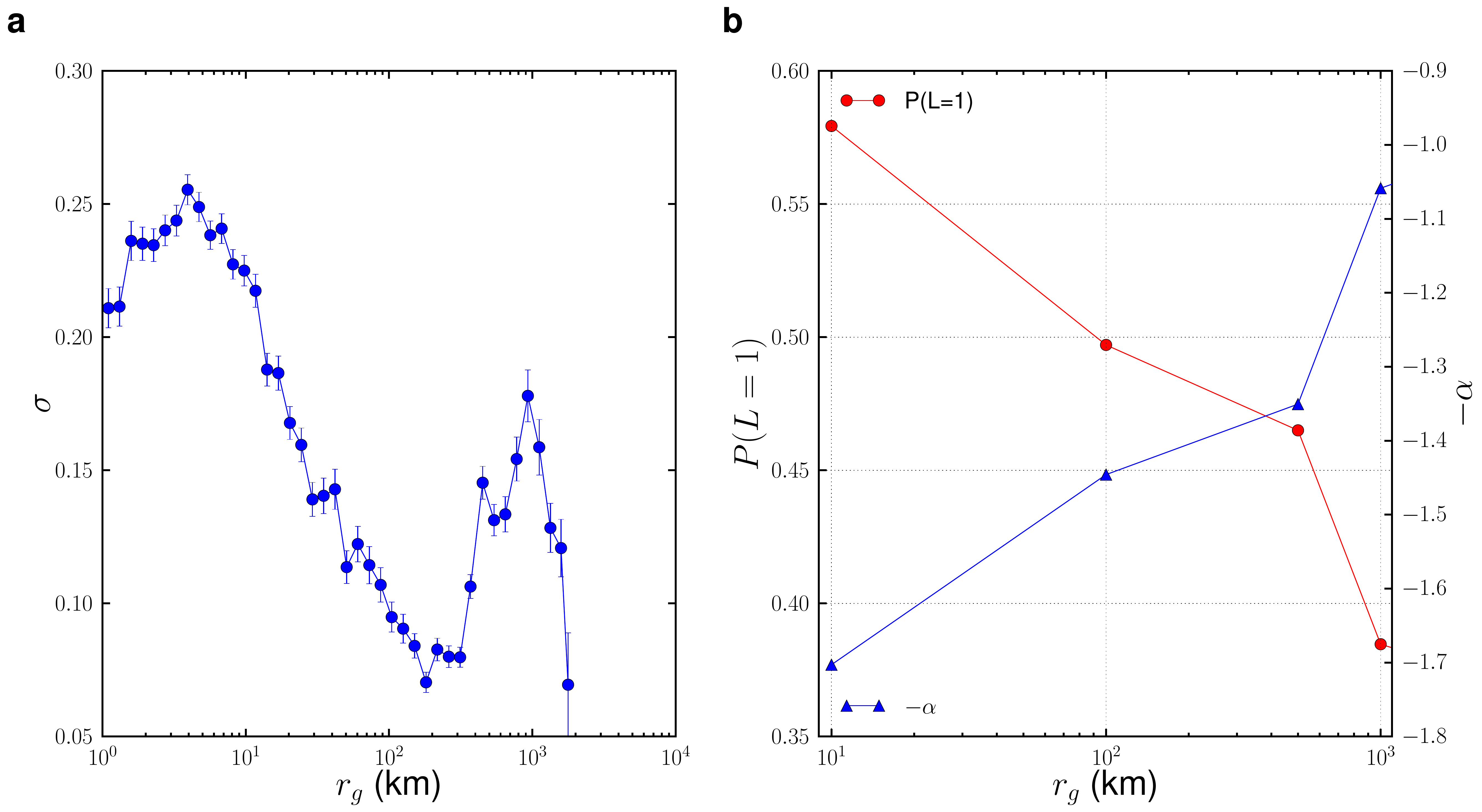}

\caption{{\bf (a)The isotropy ratio $\sigma$ steadily decreases with $r_g$ before increasing again from 200km to around 1000km, indicating that popular intercity trips contribute to increased isotropy.} (b) The probability of return to the most popular location and more generally the preference for previously visited locations both drop significantly with increasing $r_g$.}
\label{fig:4}
\end{figure}

Next, we analyse the capacity of geotagged Tweets to capture the spatial orbit of movement for groups of the population defined according to their radius of gyration $r_g$. We explore the probability density function $P(x, y)$, i.e. the probability that a user is observed at location $(x,y)$ in its intrinsic reference frame (see \cite{Gonzalez2008} for details). We measure this function for user groups of different radius of gyration $r_g$, as shown in Figure \ref{fig:Pxy}. We use the isotropy ratio~\cite{Gonzalez2008} $\sigma=\delta_y/\delta_x$, where $\delta_y$ is the standard deviation of $P(x,y)$ along the y-axis and $\delta_x$ is the standard deviation of $P(x,y)$ along the x-axis, to characterise the orbit of each $r_g$ group. At very short $r_g$ up to 4km, the isotropy ratio slightly increases (see Figure~\ref{fig:4}(a)). As $r_g$ increases furthers, we observe an increase in anisotropy in $P(x,y)$ as in~\cite{Gonzalez2008}; however, this correlation between increased anisotropy and $r_g$ is only valid for shorter distances between 4km to 200km, which maps well to typical distances for the use of cars as a transport mode. Movement patterns become more diffusive (isotropic) once again for $r_g$ between 200km and 1000km. In fact, we observe an unexpected steady rise in $\sigma$ for distances between 200km and 1000km, where people typically consider modes of transport other than cars, such as trains or planes. The peak in $\sigma$ is most likely a product of the population distribution in Australia. The top 3 cities account for more than half the population, and the distances between the largest city and commercial capital (Sydney) and the next two cities (Melbourne and Brisbane) are around 963km and 1010km respectively. This result suggests that frequent travellers among these cities are less directed and more diffusive in their movement within the cities, thus the higher isotropic ratio. 

\begin{figure}[h]
\centering
\includegraphics[width=\columnwidth]{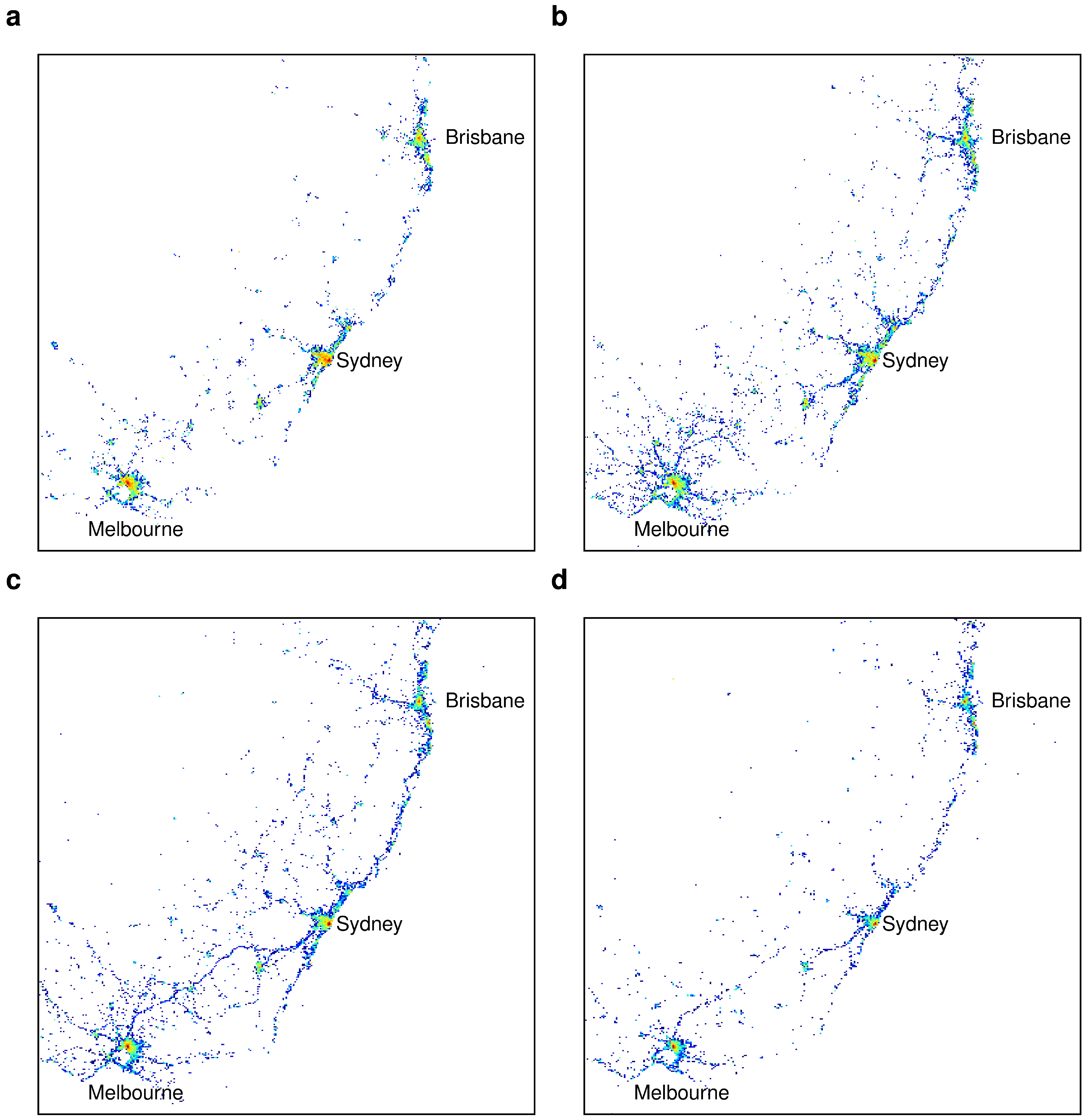}
\caption{{\bf Differences in tweet spatial distributions as the radius of gyration varies: (a) 1-10 km; (b) 10-100 km; (c) 100-500 km; (d) $>$500 km.} The maps focus on the southeast region in Australia that accounts for nearly half the population. Tweet activity for $1<r_g<10$ and $500<r_g<1000$ is mainly concentrated in large cities, while tweets for intermediate $r_g$ extend further along main highways and other regions between cities.}
\label{fig:tweet_distribution}
\end{figure}

Longer distance movers in Figures~\ref{fig:Pxy}(c) and (d) appear to have a more stretched component in the negative x-axis, with an expanding gap close to the origin as $r_g$ increases.  To shed further light on this effect, Figure~\ref{fig:tweet_distribution} compares the spatial distribution of tweets for the four categories of $r_g$, focusing on the southeast region that includes more than half the country's population (see Supporting Information S1 Text for full maps of Australia). It shows that for smaller $r_g$, tweets are concentrated in clusters mainly in large cities, or other regional areas. For intermediate $r_g$, we observe a much stronger tendency of tweets to be within an expanded region around key cities and along main roads connecting large cities. The tight coupling of movement at these distances with road usage explains the increased directivity (and anisotropy) of motion for these $r_g$ categories. The tweet distribution for large $r_g>500km$ shows a completely different trend, with a renewed focus of tweet activity in and closely around the main cities. This difference likely stems from the change in mode of travel to airplanes, where people fly in to a destination with the intention of remaining within a limited orbit around this destination. These long distance movers  appear to have a few target destinations, such as airports at key cities or locations of interest.  As we average the movement patterns over a population, the dominant movement distances for each individual may vary widely, contributing the stretch of this tail along the negative x-axis in Figures~\ref{fig:Pxy}(c) and (d). The gap that appears on the negative x-axis close to the origin arises from the use of air travel, where people do not tweet between source and destination as they travel long-distances. These patterns may also be related to the sparse and concentrated population in Australia with heavy concentration of people (and likely Twitter users~\cite{Frank2013}) at  major population centres. Another likely cause of this pattern is the fly-in/fly-out worker phenomenon~\cite{Storey2001}, where workers in the mining sector stay at remote sites during weekdays and then return home or travel to holiday destinations in Southeast Asia during weekends.  It is also likely that the increased isotropy for distances from 200km up to 1000km is due to long distance movers circulating in the vicinity of their destination away from home, given the high cost~\cite{Yan2013} associated with returning to their home location. These long distance travellers may be sending tweet messages mostly from their main destination (such as the arrival airport or a remote work site), and less frequently tweeting from satellite locations.  

Given this disparity in movement patterns for different $r_g$, we revisit the preferential return to previously visited locations for people with different $r_g$. We specifically focus on the probability of return $P(L=1)$ to the most popular (home location) and the exponent $\alpha$ of the Zipf's law fit. The reader is pointed  to Supporting Information S1 Text for full details of the analysis. We observe that the preferential return for the top location $P(L=1)$ steadily decreases with increasing $r_g$ (Figure~\ref{fig:4}(b)). The results indicate that local movers are nearly 50\% more likely to tweet from their preferred location as long distance movers, with $P(L=1)$ ranging from up to 0.8 for $1km<r_g<10km$ to as low as 0.64 for $500km<r_g<1000km$. The slope of the preferential return curve $\alpha$ is negatively correlated with $P(L=1)$. We observe a monotonic decrease in $\alpha$ for increasing $r_g$, further highlighting that preferential return weakens with longer $r_g$, not only for the most preferred location $P(L=1)$, but also for the general set of visited locations.  The likely reason for the weaker preferential return for long distance movers is the high social cost they incur for moving, which reduces the perceived value of frequently returning to specific previously visited locations.

\section*{Discussion}

We have found that human mobility patterns extracted from geotagged tweets have similar overall features as observed in mobile phone records, which demonstrates that Twitter is a suitable proxy for studying human mobility. However, marked differences are clear for tweet-based mobility data compared to other modalities. First, the higher resolution of Twitter data has uncovered a heterogeneous distribution of displacement and radius of gyration that appear to map to different modes of movement, namely intra-site, metropolitan, and inter-city movement.  Secondly, we have identified two types of Twitter users in terms of the predictability of tweeting locations, a group that is highly persistent and predicable in their tweet location probably as a consequence of how the use the technology, and another group that is much more diverse and less predictable in their tweet locations. It is for the latter group that we conjecture Twitter captures representative mobility patterns of users given the reduced preference for tweeting from a few select locations. It would be interesting to further study the effect of tweeting behaviour on the subsampling of real mobility pattern using a secondary source that provides a ground-truth for a continuously tracking of the real mobility. Thirdly, Twitter data reveals unexpected dynamics in mobility particularly for long distance movers, who are more diffusive in their movement than intermediate distance movers, most likely as a reflection of a switch in transportation mode towards air travel and local circulation around destination cities. We have also found that the likelihood of tweeting from the home location and more generally the strength of preferential return are strongly dependent on a person's orbit of movement, with long distance movers less likely to return to previously visited locations.  Our results indicate that strict limit on tweet content does not appear to change the daily cycle in returning to the same location or the preferential return trends compared to call data records. Overall, we find that population-level mobility patterns are well-represented by geo-tagged tweets, while individual-level patterns are more sensitive to contextual factors, such as the individual's degree of preference for tweeting from one or a few locations. 

Our findings can be used for improved modelling of human movement and for better characterisation of Twitter-based mobility patterns. For instance, epidemiologists modelling the risk of disease spread across a landscape can use our findings to create user movement profiles based on $r_g$, where long distance movers tend to stay in and around big cities. This may narrow down the population of likely  vectors for diseases that emerge in rural areas. Our observations on preferential return can also be used for more fine-grained modelling of individual movement based on the user profile, as suggested in~\cite{Yan2013}. These individual-based models can feed into not only disease spread forecasting, but also into the planning of communication and transportation networks. 

Another implication of our work is for further studies in geography and demography. Greater understanding of the mobility patterns of geotagged tweets can be used by geographers and demographers to model human movement and to understand the underlying drivers for people moving~\cite{Bernard14}. Coupled with tweet contents, these mobility dynamics can provide a useful tool for new methodologies in human geography studies, particularly for population projection. Despite mobility being the main driver of population changes, current population projection methods rely on coarse-grained census data. Twitter-based tracking provides high spatiotemporal data for more realistic mobility assumption for population projection models.

The current study is limited to geotagged tweets, which only account for a small portion of tweets. To fully exploit the potential of Twitter in human mobility analysis, an interesting direction for future work is to apply the similar approach to tweets without geotags using location inference~\cite{Ikawa12} based on the tweet context. The insight from this article on mobility dynamics from tweets also lays the groundwork to generate synthetic mobility data~\cite{Isaacman2012} at various spatial scales for analysis of disease spread, transport systems or communication networks.

\section*{Methods}
\subsection*{Dataset}
Data have been collected through the public Twitter Stream API (https://dev.twitter.com/overview/api), as part of the Emergency Situation Awareness project~\cite{Cameron2012}.Our dataset consists of $7,811,004$ tweets from $156,607$  different users in Australia from Sep. 1 2013 to March 31 2014. In this dataset, each tweet record has a geotag and a timestamp indicating where and when the tweet was posted. Based on this information we are able to construct a user's location history denoted by a sequence $L=\{(x_i, y_i, t_i)\}$. The original location information provided by the geotag is denoted by latitude and longitude, and for convenience we project the locations in the EPSG:3112 coordinate system. We first filter the dataset to exclude tweets that are not posted within Australia, restricting our study in the domestic domain. We then filter users who have unrealistic moving patterns to reduce the noise and outliers of the dataset. In this study, if the displacement between two consecutive locations is not traveled at an usual velocity, i.e. $\frac{d}{\Delta T} > v_c = 240m/s$, we consider the mobility pattern is unrealistic. The following study is then based on the filtered dataset which contains 4,171,225 tweets and 79,055 users.\\

\subsection*{Identification of locations}
To measure the visitation frequency, we first need to identify locations. Due to the spatial uncertainty, proximate locations in the raw data can represent an identical location of interest. Therefore we use DBSCAN clustering~\cite{Ester1996} to eliminate the vagueness, i.e. locations in the same cluster are considered to be one single location. The advantage of DBSCAN is that it can identify clusters of arbitrary shape. In particular, we use $\epsilon=250m$ and $n_{min}=1$ in the DBSCAN clustering which represent the threshold distance and the minimum number of points to form a cluster respectively. The effect of changing the cluster size is investigated in Supporting Information S1 Text.

\subsection*{Entropy and predictability}
Let the sequence of tweeting locations of user $i$ be $T_i=\{X_1,X_2....X_L\}$, where each symbol $X_k \in \{1, \dots, N_i\}$ denotes the user's $k^{th}$ tweeting location and $N_i$ is the number of distinct visited locations of user $i$ during the observation period.

The unconditional entropy $S^i_{unc}$ (or Shannon entropy) of $T_i$ can be expressed as:
\begin{equation}
S^i_{unc}=- \sum_{j=1}^{N_i}{p_i (j)\log_2 p_i(j)}
\end{equation}
Where $p_i(j)$ is the historical probability that location $j$ was visited by the user $i$.  

The real entropy $S^i_{real}$ can be expressed as follows~\cite{Song2010_2}:
\begin{equation}
S^i_{real}=-\sum_{T_i'\subset{T_i}}{P(T_i')log_2[P(T_i')]}
\end{equation}
where
$P(T_i')$ is the probability of finding a particular
time-ordered subsequence $T_i'$ in the trajectory
$T_i$. The Shannon entropy is inherently higher than the real entropy for users with sufficient tweet history~\cite{Song2010_2}. The real entropy can be estimated using a Lempel-Ziv (LZ) algorithm that searches for repeated sequences.

Based on Fano's inequality~\cite{Fano1961}, the maximum bound of the user's predictability $\Pi\leq\Pi^{max}$ can be obtained from the entropy via the following equation :
\begin{equation}
S=H(\Pi^{max})+(1-\Pi^{max})log_2(N-1)
\end{equation}
The binary entropy function $H$ is given by:
\begin{equation}
H=-\Pi^{max}log_2(\Pi^{max})-(1-\Pi^{max})log_2(1-\Pi^{max}).
\end{equation}

\section*{Supporting Information}
\subsection*{S1 Text}
\label{S1_text}

\section*{Acknowledgments}

\section*{Author contributions}
RJ and KZ wrote the main manuscript text. All authors conducted data analysis and results synthesis. All authors reviewed the manuscript.

%
%
%

\end{document}